# Teaching Galileo? Get to know Riccioli!

# What a forgotten Italian astronomer can teach students about how science works.

Christopher M. Graney

Jefferson Community & Technical College
1000 Community College Drive
Louisville KY 40272 (USA)
christopher.graney@kctcs.edu

What can physics students learn about science from those scientists who got the answers wrong?  Your students probably have encountered little science history.  What they have encountered probably has portrayed scientists as The People with the Right Answers.  But those who got the wrong answers can teach students that in science answers are often elusive -- not found in the back of a book or discovered in a bold stroke of genius.

The bold successes of Einstein are standard lore in the broader culture.  But who outside of the world of physics has heard of Michaelson and Morely -- very good physicists who helped set the stage for Einstein with their unsuccessful experiment to detect Earth's motion through the "Ether" (the supposed medium for light waves)?  Even greater lore is attached to Galileo -- the "Father of Physics".  Einstein characterized Galileo as being a "representative of rational thinking" who led humanity to an objective and causal attitude toward the cosmos by standing up against a host of those who simply defended authority.[1]  In the broader culture, Galileo's work with the telescope is ranked with Columbus's voyages and Gutenberg's press.[2]  But who, even within the world of physics, has heard of one of Galileo's most ardent critics, the 17th century Italian astronomer Giovanni Battista Riccioli?  Riccioli (1598-1671) is known for his map of the moon (Figure 1) which established our modern system of lunar nomenclature.  He is credited with the first measurement of the acceleration due to gravity *g,* which he measured to be 9.6 m/s$^2$ (a value he obtained, through remarkable dedication to experimental accuracy, using merely a pendulum and the stars to time falling bodies)[3].  And he is also known for his discussion of Heliocentrism vs. Geocentricism.

In this discussion, "the lengthiest, most penetrating, and authoritative analysis made by any author of the sixteenth and seventeenth centuries",[4] Riccioli laid out 77 arguments against the heliocentric hypothesis of Copernicus that a person in the 17th century might have encountered.[5]  Riccioli said that most of these anti-Copernican arguments were without merit.  For example, an argument that said Earth cannot be moving because the speed of its rotation would overwhelm the flight of birds and the movement of ships[6] was easily answered by appeal to the idea of common motion (which was how Galileo answered such arguments in his *Dialogue Concerning the Two Chief World Systems*[7]).  Likewise, an argument that a moving Earth would destroy the absolute sense of downward movement by heavy objects found in a geocentric universe[8] was easily answered by saying that the downward motion of heavy bodies was toward the center of the Earth, not the center of the Universe.

---

[1] Albert Einstein, "Foreword" (p. xxiii) in Galileo Galilei, *Dialogue Concerning the Two Chief World Systems, Ptolemaic and Copernican*, translated by Stillman Drake (Random House, New York, 2001).

[2] *The LIFE Millenium: The 100 Most Important Events & People of the Past 1,000 Years*, Robert Friedman, ed. (Life Books, New York, 1998).  LIFE ranked only Gutenberg, Columbus, Luther, and the Industrial Revolution ahead of Galileo.

[3] Riccioli created a pendulum with a precise half-period of one second by timing pendulums via the stars and very, very long sessions of counting swings (boosting a pendulum when its amplitude decreased too much).  These included at least one 24-hour marathon session involving a team of counters (monks from Riccioli's order -- he was a Jesuit priest).  Pendulums with shorter periods were calibrated from the seconds pendulum.  To time balls falling from the 100 meter tall Torre degli Asinelli in Bologna, Italy, he recruited a chorus of monks to chant notes to the rhythm of the faster pendulums, creating an audible "timer" which could determine fall times to a high degree of accuracy.  See J. L. Heilbron, *The Sun in the Church: Cathedrals as Solar Observatories* (Harvard University Press, 2001), pp. 180-81.

However, Riccioli said not all the 77 arguments were so easily dismissed.  Several were based on the idea that a rotating Earth would deflect falling bodies and projectiles -- what today we call the "Coriolis Effect" (Figure 2).[9]  Others were rooted in observations of stars.  Astronomers of the time did not understand how light from a point source diffracts through a circular aperture to form a spurious Airy Disk -- they thought the disks they saw when looking at stars through a telescope (Figure 3) were the stars themselves.[10] And, at the distances the stars had to lie at for the motion of the Earth to produce no yearly changes in the appearance of stars (an effect known as annual parallax), these disks would translate into immense stars, vastly larger than the sun, perhaps larger than the Earth's whole orbit or even than the entire solar system![11]

---

[4] Edward Grant, *Planets, Stars, and Orbs: The Medieval Cosmos, 1200-1687* (Cambridge University Press, 1996), p. 652.

[5] Riccioli's map of the moon, discussion of falling bodies, and 77 arguments are all found in his *Almagestum Novum* (Bologna, 1651), which is available on-line at http://www.e-rara.ch/zut/content/pageview/140188.  It is very interesting to look at, even if you do not know Latin!  An English summary of the 77 arguments is available on the Physics Arxiv (C. M. Graney, "Giovanni Battista Riccioli's Seventy-Seven Arguments Against the Motion of the Earth", arXiv:1011.3778, http://arxiv.org/abs/1011.3778).  All the references to Riccioli in this paper are from these sources.

[6] Argument #27 of the 77.

[7] See *Dialogue*, pp. 216-18.

[8] Argument #50 of the 77.

[9] "Forces and Fate", *New Scientist*, 8 January 2011, p. 6.

And so Riccioli said that the weight of argument favored a "geo-heliocentric" hypothesis such as that advocated by the great Danish astronomer Tycho Brahe (Figure 4). In Brahe's hypothesis the Earth is immobile while the sun, moon, and stars circle it. This agrees with the apparent absence of "Coriolis" effects. It does not require stars be distant (and apparently immense) to explain away the absence of annual parallax. But in Brahe's hypothesis, which was popular into the late 17th century and beyond (Figure 5),[12] the planets circle the sun. This agrees with telescopic observations such as the phases of Venus discovered by Galileo. It also avoids issues of physics. Prior to Newtonian ideas, there was no solid explanation for how the heavy

---

[10] Because Galileo remarked in his *Starry Messenger* that stars seen through the telescope appear much the same as when seen by the naked eye, it is often said that Galileo and other astronomers understood stars to be dimensionless points of light. However, in his writings after the *Starry Messenger* Galileo consistently said stars viewed with the telescope appeared as disks or spheres. See C. M. Graney, "Is Magnification Consistent?", *The Physics Teacher* 48, 475-477 (October 2010).

[11] C. M. Graney, "The Telescope Against Copernicus: Star observations by Riccioli supporting a geocentric universe", *Journal for the History of Astronomy* 41, 453-467 (2010).

[12] Christine Schofield, "The Tychonic and Semi-Tychonic World Systems", in *Planetary Astronomy from the Renaissance to the Rise of Astrophysics*, edited by R. Taton and C. Wilson (Cambridge University Press, 1989), Part A, p. 39.

Earth might be moved around the sun (the motion of heavenly bodies around Earth was explained by postulating that they were made of material with special properties not found on Earth, much as today astronomers postulate "dark matter" and "dark energy" to explain observations). There were also issues of religious belief -- various passages from the Christian Bible spoke of the Earth as being fixed in place and of the Sun as moving.[13] Thus Tycho Brahe had said that Copernicus's model was elegant, but it violated physics because it gave the Earth, "that hulking, lazy body, unfit for motion", a rapid, complex motion; whereas the geo-heliocentric hypothesis "offended neither the principles of physics nor Holy Scripture".[14]

Riccioli's preference for Tycho's model illustrates something important about how science is done. While today anti-Copernicans are often portrayed as Einstein characterized them (opposed to rational thinking; opposed to science), Riccioli, perhaps the most prominent of the anti-Copernicans, examined the available evidence diligently and rationally. The conclusion he reached was indeed wrong, but wrong because at that time neither the diffraction of light and the Airy Disk, nor the details of the Coriolis effect were understood. Riccioli's anti-Copernican arguments were so solid that they would become subjects of further investigation in physics, long after the Copernican theory had triumphed over the Tychonic theory.[15]

---

[13] Only two of the 77 anti-Copernican arguments Riccioli mentions dealt with religion, and he dismissed both.

[14] Owen Gingerich and J. R. Voelkel, "Tycho Brahe's Copernican campaign," *Journal for the History of Astronomy* **29**, 1-34 (1998), p. 24, p. 1.

[15] Direct evidence for Earth's motion would be discovered in 1728, when James Bradley detected "stellar aberration", a deflection of starlight caused by Earth's motion around the Sun, by which time Newtonian physics had provided a full theoretical framework for underpinning the Copernican model. Full understanding of both the

Riccioli illustrates that easy answers are rare in science; that science is a complex and nuanced undertaking where "the answer" can elude even good scientists.  This is an important idea to convey to physics students, who often are highly focused on "the answer".  Often this focus is to the exclusion of doing things that may not obviously lead to the answer in the back of the book (I can seldom convince my students, for example, to carefully draw a free body diagram before writing down equations).  Or, it leads to impatience in lab, as students don't view the problems naturally encountered in getting "the answer" in lab as normal and part of real science, but as signs that they are physics failures!  This important idea also reaches beyond the classroom and laboratory.  Not understanding how elusive "the answer" can be in science leads to some weird ideas that students may acquire from the broader culture.  For example, they may think that since science has difficulty determining if things we eat are harmful to our health (caffeine, saccharine), science should simply be dismissed regarding such matters because "in a few years they'll be telling us something else".[16]  Riccioli shows that answers can be

---

Coriolis effect and the Airy Disk would elude physicists until the early 19th century, more than 150 years after the *Almagestum Novum*.  Interestingly, the diffraction of light would be discovered and named by none other than Riccioli's assistant, Francesco Maria Grimaldi.

[16]Far more extreme ideas regarding science are not uncommon, especially the idea that a simple answer is known, but science is hiding it -- whether "it" is the cause of increasing autism rates (vaccines); or the reason we have not returned to the Moon (aliens -- unless it is that we never went to the Moon and NASA faked the whole thing in a studio); or the reason we don't have solar-powered cars (the technology is being suppressed by the energy companies).  The vaccine issue, and the problem of public distrust of science in general, has been a recurring and recent topic on NPR's "Science Friday" talk show -- see, for example, "Paul Offit and 'Deadly Choices'", Science Friday (January 7, 2011: http://www.sciencefriday.com/program/archives/201101075).  In all of these

elusive, but science does make progress -- we eventually figured out that Copernicus was right, and in time we will figure out food better, too.

I have introduced Riccioli to students in my physics classes (calculus-based) as an anecdote and in my introductory astronomy classes as a significant part of the discussion of the Copernican Revolution.  I am pleased with the results.  The physics students are encouraged by this example of how the answers, even to questions about a supposedly so simple matter as whether the Earth moves, are not easy.  The reactions of astronomy students, typically non-science majors, are more dramatic -- especially those students with a skeptical or even combative attitude toward science and with weird ideas acquired from the broader culture.  They are not expecting Riccioli.  He gets them to open up to science and think about the challenges scientists face in finding "the answer".

So next time you talk to students about Galileo and the right answer, also mention Riccioli and the wrong answer.  Your students will understand science better for it.

---

examples there is an implicit rejection of any sense that answers might be challenging and elusive in science; rather, the assumption is that the answer is known, but kept hidden.

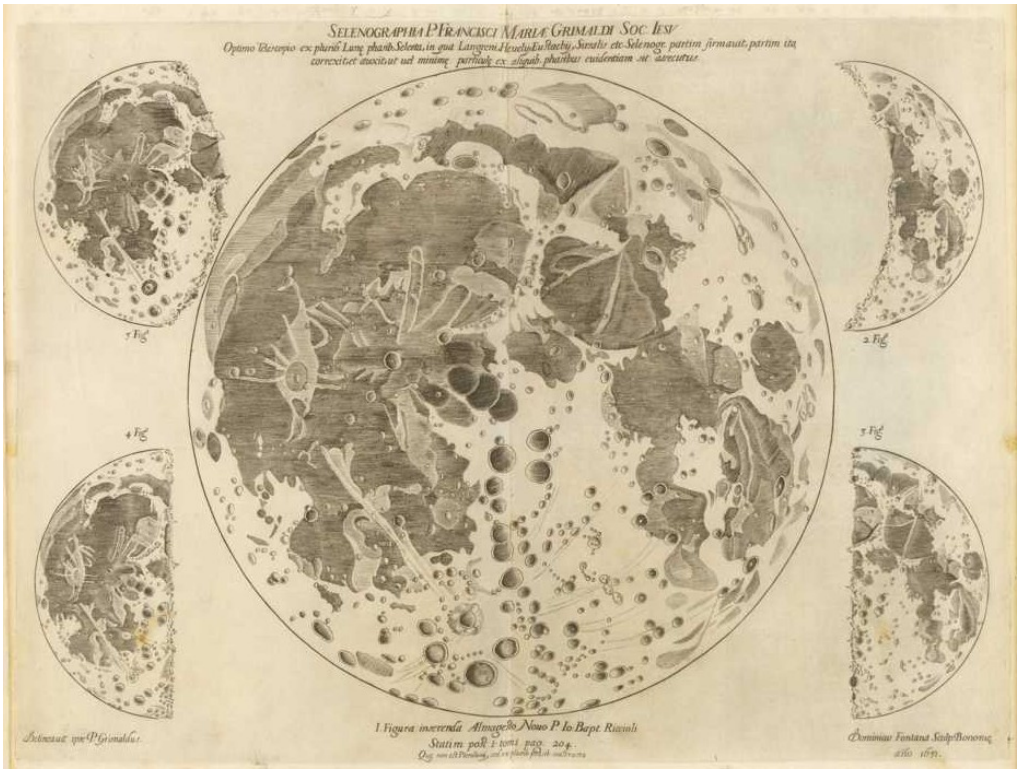

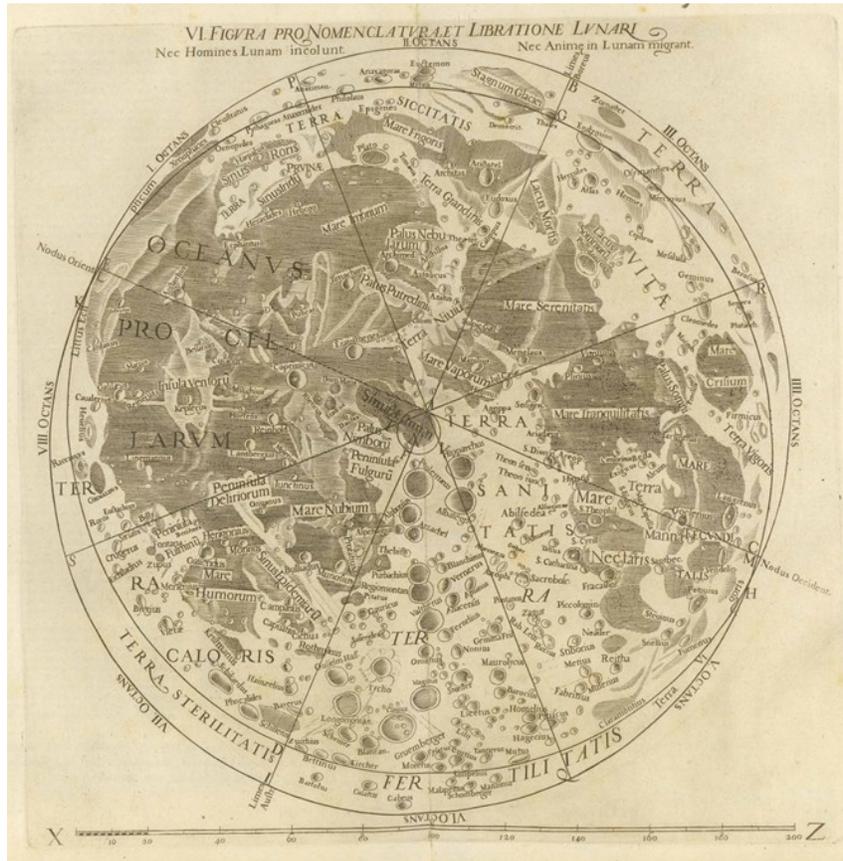

**Figure 1.**

**Maps of the Moon from Riccioli's 1651** *Almagestum Novum,* **created by Riccioli and his assistant, Francesco Maria Grimaldi (1618-63). The names of prominent lunar features, such as the Sea of Tranquility, originated with these maps.**

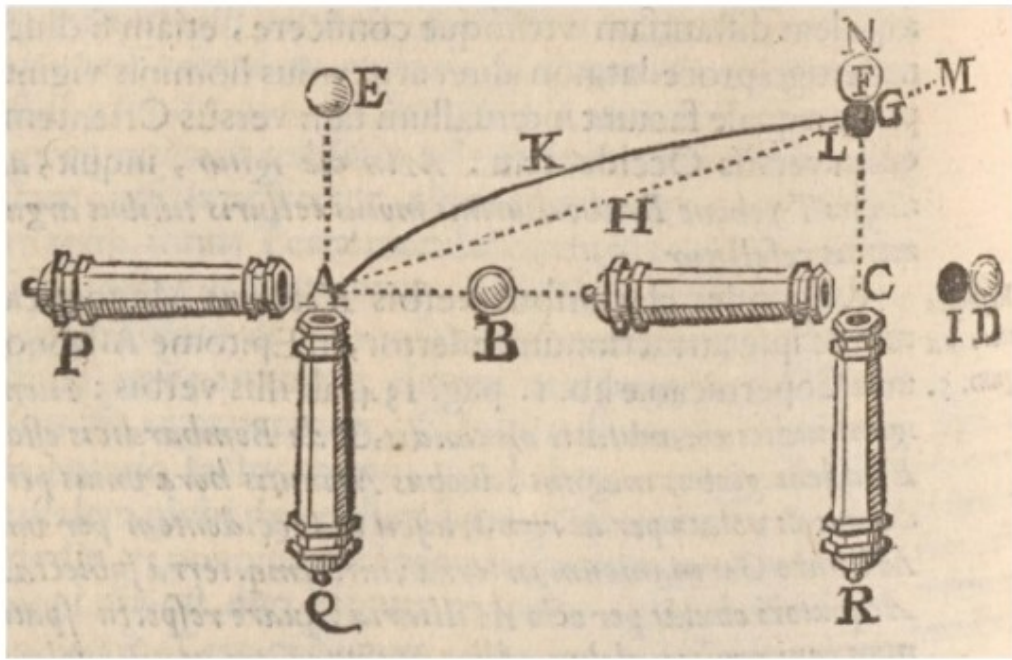

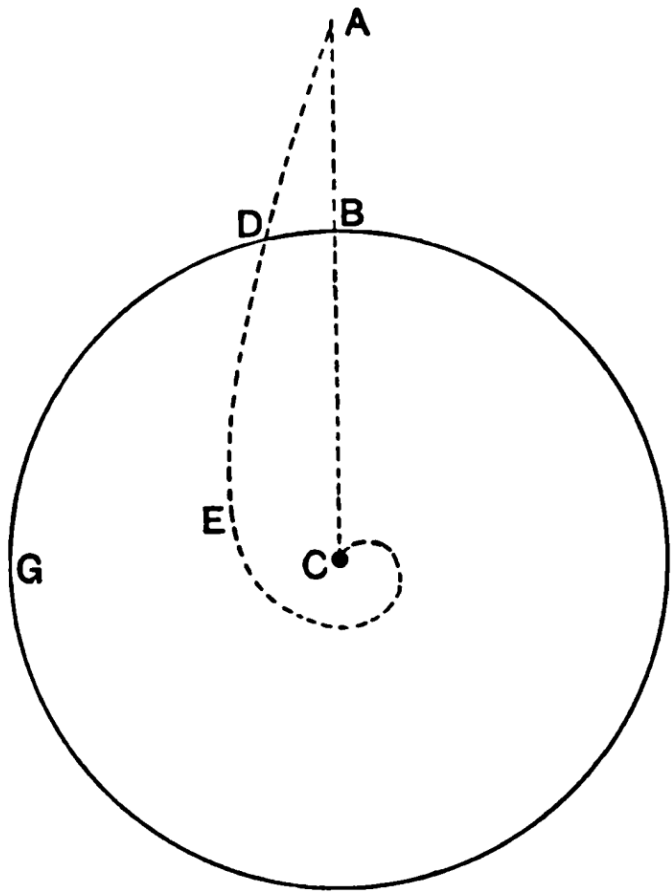

**Figure 2.**

**Top -- Diagram from the *Almagestum Novum* illustrating why a rotating Earth should cause deflection in the trajectory of a cannonball fired northward. Riccioli says that because the ball passes over slower-moving ground as it travels north, as seen from the cannon it will bend to the east, striking at G instead of at the intended point F.**

**Bottom -- Riccioli also argued that an object falling from a fixed point above the Earth would fall vertically if the Earth were immobile, but would arc to the east if Earth rotated (owing to the greater tangential speed of the object than of the point on Earth's surface directly below it). This diagram -- from Walter William Rouse Ball's 1893 *An Essay on Newton's 'Principia'*, p. 142-3 -- is actually from a 1679 letter from Isaac Newton to Robert Hooke proposing that Earth's movement could indeed be detected through this phenomenon, which prompted Hooke to try to do so (unsuccessfully).**

**Today these phenomena are recognized as the Coriolis Effect in action. Riccioli took the fact that such effects had not been observed to be evidence for Earth's immobility.**

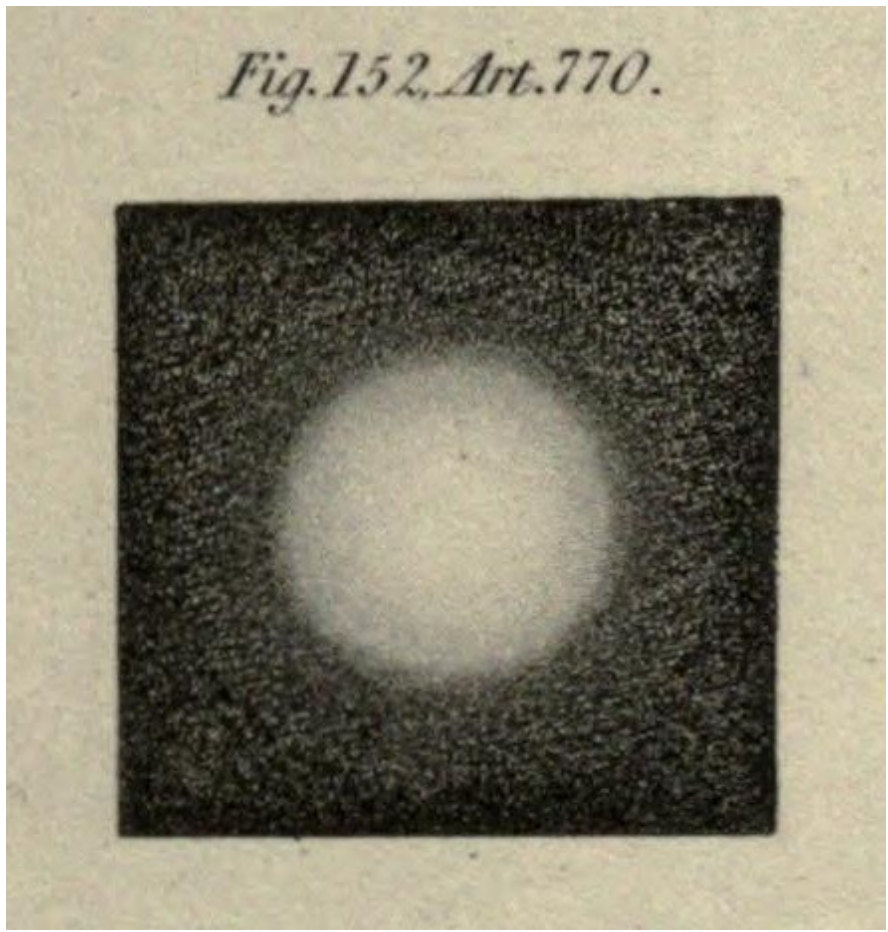

Figure 3.  Sketch of a star seen through a small aperture telescope such as was used in much of the 17[th] century (from John Herschel's article on "Light" for the 1828 *Encyclopædia Metropolitana*).  This globe-like appearance, the spurious "Airy Disk" formed by light diffracting through the telescope's aperture, was understandably interpreted by early telescopic astronomers (including Galileo) as the star's physical body.  Riccioli argued that under the Copernican hypothesis (which required stars to be extremely distant), stars must be orders of magnitude larger than even the Sun to have such an appearance -- thus the Copernican hypothesis was absurd.

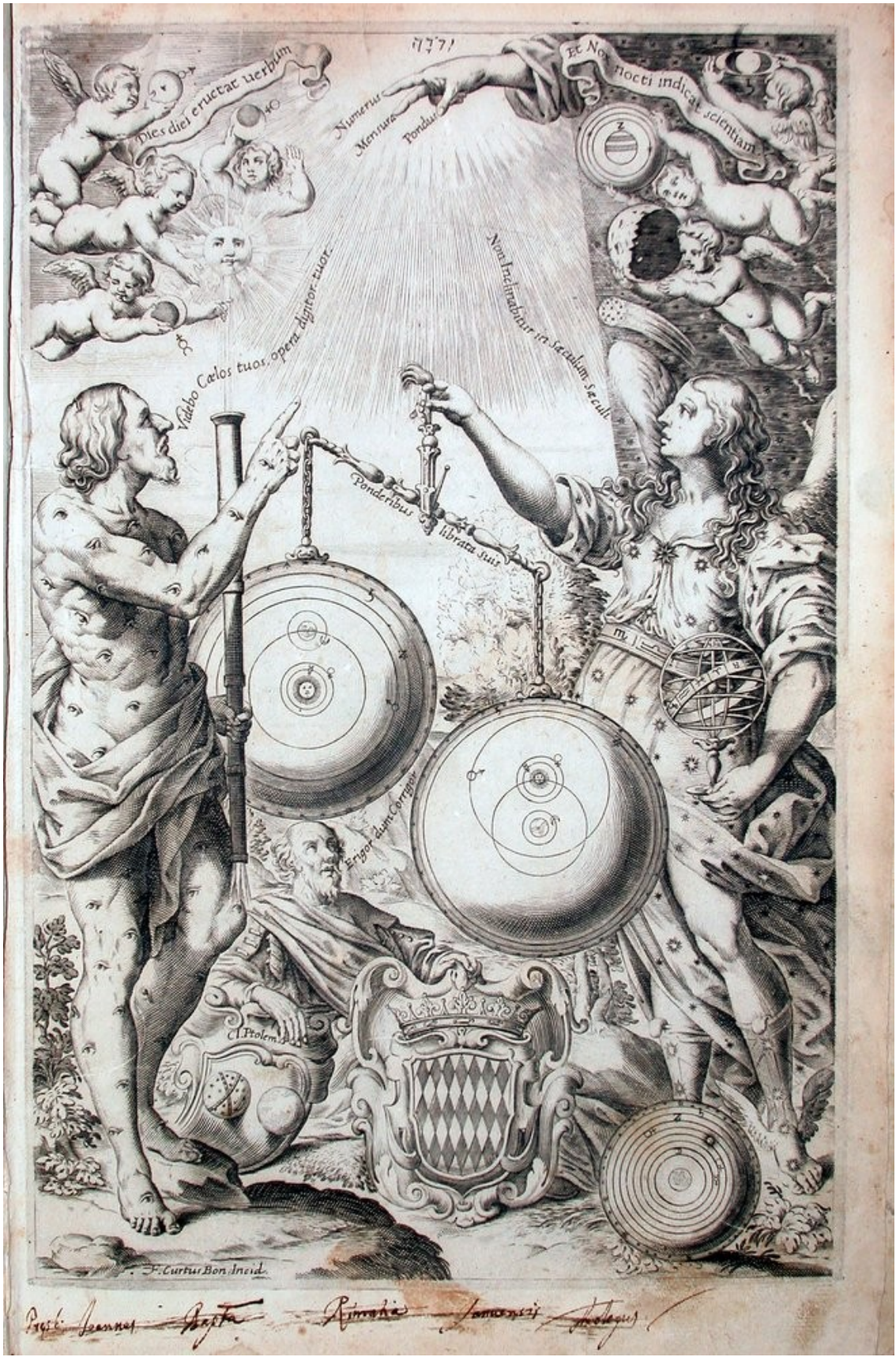

Figure 4.

Frontispiece of the *Almagestum Novum*, showing Riccioli's assessment of the debate over whether the Earth moved.  Mythological figures Argus (holding the telescope) and Urania (holding the scales) weigh the heliocentric hypothesis of Copernicus against a geo-heliocentric hypothesis such as Tycho Brahe promoted.  The old purely geocentric model, in which everything circles the Earth, lies discarded on the ground, disproven by discoveries made with the telescope.  These discoveries, which include phases of Venus and moons of Jupiter, are illustrated at top left and right.  The balance tips in favor of the geo-heliocentric hypothesis, showing Riccioli's opinion about how the debate stood at the time.

Image courtesy History of Science Collections, University of Oklahoma Libraries.

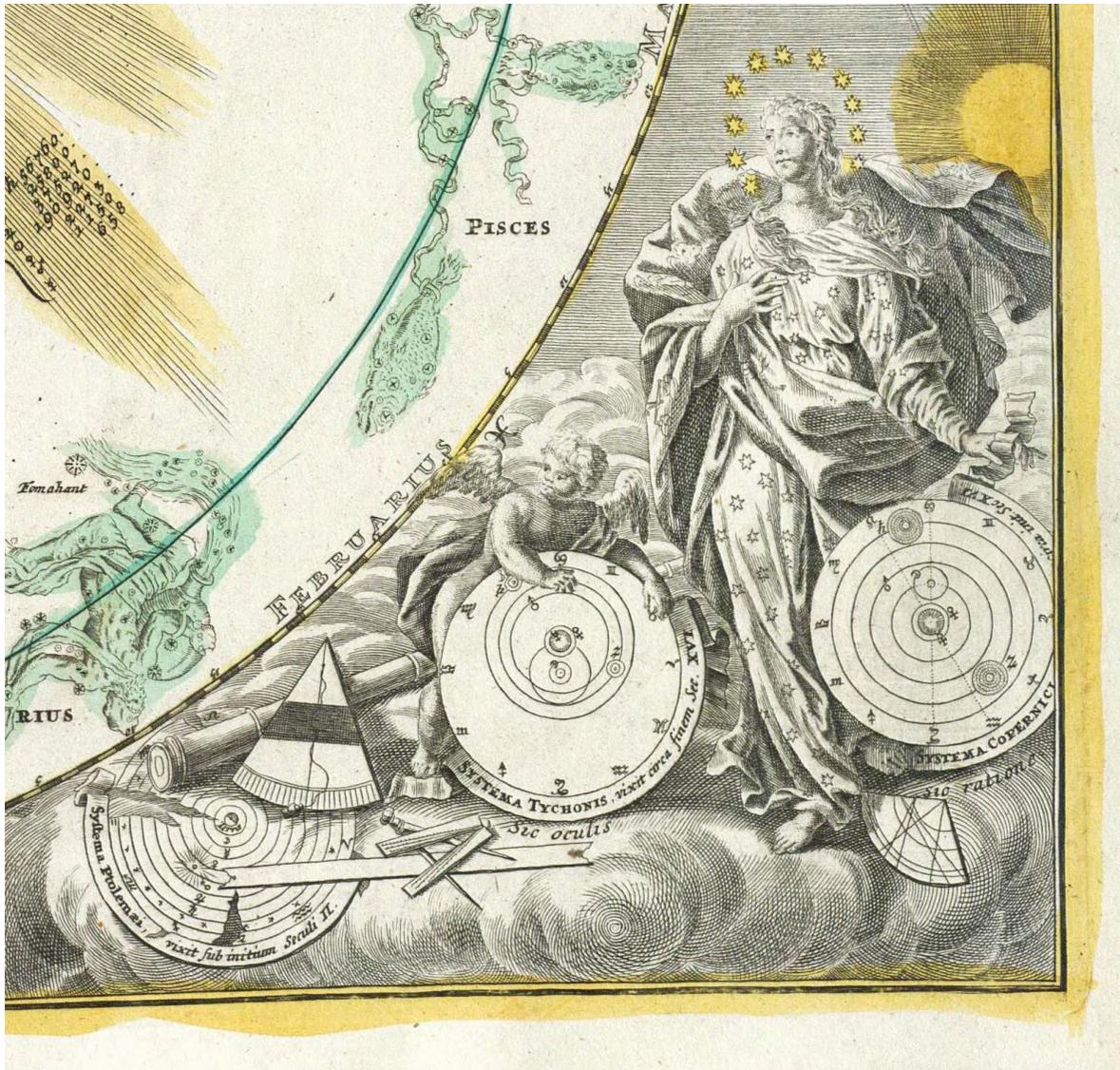

**Figure 5.**

**Illustration from the 1742 *Atlas Coelestis* by J. G. Dopplmayer and J. B. Homann. (Compare to Figure 4) The old purely geocentric model is shown as broken under the telescope and discarded; the choice for scientists is between the heliocentric and geo-heliocentric hypotheses. Here however, it is the heliocentric that is shown as being the better choice. But a significant portion of the *Atlas Coelestis* is devoted to the geo-heliocentric hypothesis. A hypothesis with an immobile Earth had staying power a full century after Galileo.**

**Image courtesy of R. H. van Gent.**